\documentclass[twocolumn]{aastex7}

\newcommand{\myr}{\; {\rm Myr}}
\newcommand{\gyr}{\; {\rm Gyr}}
\newcommand{\K}{\; {\rm K}}
\newcommand{\au}{\; {\rm au}}
\newcommand{\um}{\; {\mu {\rm m}}}
\newcommand{\mas}{\; {\rm mas}}

\begin{document}

\title{Theory of Exozodi Sources and Dust Evolution}

\author[orcid=0000-0001-9064-5598,sname=Wyatt,gname=Mark]{Mark C. Wyatt}
\affiliation{Institute of Astronomy, University of Cambridge, Madingley Road, Cambridge, CB3 0HA, UK}
\email[show]{wyatt@ast.cam.ac.uk}

\author[orcid=0000-0001-5653-5635,sname=Pearce,gname=Tim]{Tim D. Pearce}
\affiliation{Department of Physics, University of Warwick, Gibbet Hill Road, Coventry, CV4 7AL, UK}
\email[]{Tim.Pearce@warwick.ac.uk}

\author[orcid=0000-0002-9385-9820,sname=Pawellek,gname=Nicole]{Nicole Pawellek}
\affiliation{Institut f\"{u}r Astrophysik, Universit\"{a}t Wien, T\"{u}rkenschanzstrasse 17, 1180, Vienna, Austria}
\email[]{nicole.pawellek@univie.ac.at}

\author[orcid=0000-0002-8796-4974,sname=Dodson-Robinson,gname=Sarah]{Sarah Dodson-Robinson}
\affiliation{University of Delaware, 210 South College Ave, Newark, DE 19716, USA}
\email[]{sdr@udel.edu}

\author[orcid=0000-0001-6403-841X,sname=Faramaz,gname=Virginie]{Virginie C. Faramaz-Gorka}
\affiliation{Department of Astronomy and Steward Observatory, University of Arizona, 933 N Cherry Ave., Tucson, AZ 85721-0065, USA}
\email[]{vfaramaz@lbti.org}

\author[orcid=0000-0002-4388-6417,sname=Rebollido Vazquez,gname=Isabel]{Isabel Rebollido}
\affiliation{European Space Agency (ESA), European Space Astronomy Centre (ESAC), Camino Bajo del Castillo s/n,
28692 Villanueva de la Ca\~{n}ada, Madrid, Spain}  
\email[]{Isabel.RebollidoVazquez@esa.int}

\author[orcid=0000-0003-4842-5721,sname=Rigley,gname=Jessica]{Jessica K. Rigley}
\affiliation{Institute of Astronomy, University of Cambridge, Madingley Road, Cambridge, CB3 0HA, UK}
\email[]{jessica.rigley@cantab.net}

\author[orcid=0000-0000-0000-0000,sname=Stark,gname=Christopher]{Christopher C. Stark}
\affiliation{NASA Goddard Space Flight Center, Exoplanets and Stellar Astrophysics Laboratory, Code 667, Greenbelt, MD 20771, USA}
\email[]{christopher.c.stark@nasa.gov}

\begin{abstract}

Exozodiacal dust disks (exozodis) are populations of warm ($\sim 300\K$) or hot ($\sim 1000\K$) dust,
located in or interior to a star's habitable zone, detected around $\sim 25$\% of main-sequence
stars as excess emission over the stellar photosphere at mid- or near-infrared wavelengths.
Often too plentiful to be explained by an \textit{in-situ} planetesimal belt,
exozodi dust is usually thought to be transported inwards from further out in the system.
There is no consensus on which (if any) of various proposed dynamical models is correct,
yet it is vital to understand exozodis given the risk they pose to direct imaging
and characterisation of Earth-like planets.
This article reviews current theoretical understanding of the origin and evolution of exozodi dust.
It also identifies key questions pertinent to the potential for exozodis to impact
exoplanet imaging and summarises current understanding of the answer to them informed by exozodi theory.
These address how exozodi dust is delivered, its size and spatial distribution, and the effect
of its composition on exozodi observability, as well as the connection between hot and warm exozodis.
Also addressed are how common different exozodi levels are and how that level can be predicted from system properties,
as well as the features that planets impart in dust distributions and how exozodis affect a planet's physical
properties and habitability.
We conclude that exozodis present both a problem and an opportunity, e.g., by introducing noise that makes
planets harder to detect, but also identifying systems in which ingredients conducive to life, like water and volatiles,
are delivered to the habitable zone.

\end{abstract}

\section{Introduction}

\begin{figure*}[]
    \centering
    \includegraphics[width=18cm]{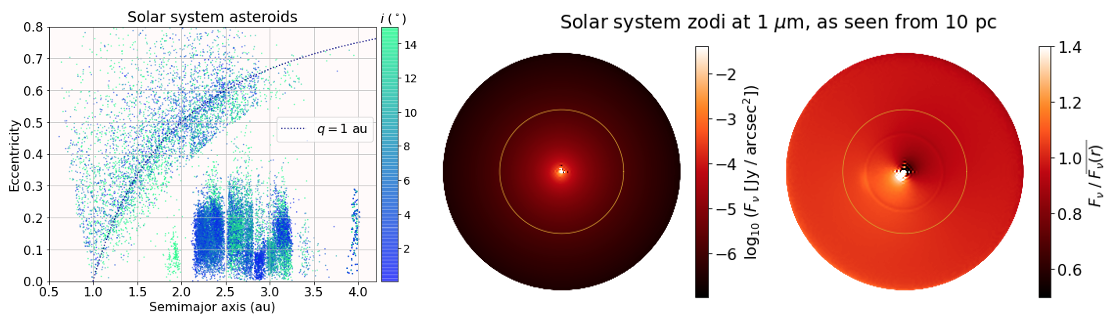}
    \caption{Distribution of asteroids and dust in the inner Solar system.
    (Left) Orbital distribution (eccentricity-semimajor axis) of 20,000 asteroids,
    colour-coded by inclination (orbits retrieved from the Minor Planet Center).
    The Main Belt asteroids are concentrated $2.1-3.3$\,au, while Near-Earth asteroids are found in the region between
    where pericentres ($q=a(1-e) \approx 1$\,au) and apocentres ($Q=a(1+e) \approx 1$\,au).
    (Right) Model images of the zodiacal light at 1\,$\micron$ as seen from 10\,pc (created using {\it zodipic}).
    The middle panel shows surface brightness, while on the right this is normalised by the average surface brightness
    at each radius to accentuate the Earth’s resonant ring (at $\sim 1$\,au) and the asymmetry caused by Jupiter.
    The yellow circles mark a distance of 1.5\,au.}
    \label{fig:sf1}
\end{figure*}

Planetary systems are made up of multiple components.
There are the planets themselves, which may be analogues to the terrestrial, gas giant or ice giant planets in the Solar
system, but may be more exotic, such as the common class of super-Earth or sub-Neptune exoplanets found orbiting
$\ll 1$\,au from $\sim 30$\% of Sun-like stars \citep[e.g.,][]{Winn2015}.
Then there are the smaller bodies, like asteroids and comets (collectively termed planetesimals), which may lie in asteroid
belt, Kuiper belt or Oort cloud analogues, but may again have more exotic architectures.
These smaller bodies extend in size down to $\micron$-sized dust, which is thought to have its source in the planetesimal
populations, but which can pervade the system extending far from its source.
For example, the Earth is embedded in a disk of dust known as the zodiacal cloud (see Fig.~\ref{fig:sf1}), which originates
in the collisional grinding of asteroids and the disintegration of comets at a few au \citep{Nesvorny2010}.
Infrared and sub-mm observations probing $\micron$-cm-sized dust show that $\sim 20$\% of stars have a cold
$>10$\,au planetesimal belt \citep{Wyatt2008}, but we are only just beginning to understand the exact sizes of the parent
planetesimals and their dynamical relation to any planets in their systems.
The focus of this article is the population of dust around other stars that is found in or interior to its habitable zone,
i.e., within a few au for a Sun-like star \citep{Kral2017}. 
By analogy with the zodiacal cloud this dust is collectively known as a star's exozodiacal cloud, or exozodi, and
depending on its temperature it is often referred to as warm ($\sim 300$\,K) or hot ($\sim 1000$\,K) dust.

As with all of the components of a planetary system, the study of exozodis can be uniquely informative about the architecture
and evolutionary history of that system.
The dust distribution can provide evidence pointing to other components (such as planets and planetesimals) which may be present
but hard to detect, and imprinted within it can be the signature of past or present dynamical interactions with planets
\citep[e.g.,][see also Fig.~\ref{fig:sf1}]{Currie2023}.
The study of exozodis also has a particular motivation due to the colocation of the dust with the star's habitable zone.
That is, exozodis probe the environment within which potentially habitable planets orbit, and so the material which may
interact with and influence such planets \citep[e.g.,][]{Walton2024}.
They also present emission from the same vicinity as these potentially habitable planets.
This latter point means that exozodis can be problematic, since they can confound attempts to detect the light from
habitable planets \citep{Roberge2012}.
Such attempts are a major focus of current astronomy, with missions being designed with goals of not only
finding exo-Earths \citep[e.g.,][]{Rauer2024}, but also characterising their atmospheres to search for signs of life
\citep[e.g.,][]{Alei2024}.

The aim of this article is to summarise ongoing attempts to explain exozodi observations with theoretical models.
Exozodi models have had various degrees of success, but there is no single, complete model that explains the phenomenon.
Instead, a variety of explanations have been put forward for how exozodis form and evolve.
For example, several mechanisms have been suggested to supply warm exozodis, including replenishment by activity or
disintegration of comets, or by collisions between planetesimals.
Similarly, various theories have been proposed to explain hot exozodis, however such hot dust has proven
particularly difficult to explain.
To properly assess how exozodis could impact exoplanet-imaging surveys, it is important to have comprehensive,
predictive models for how this dust is distributed, how much there is, where it comes from and how it evolves.
Such an understanding would allow an assessment of which systems would be favourable for exoplanet imaging, and could mitigate
the effect of dust in those observations.
In order to progress, it is important to recognise not only where current models are successful, but also where they
struggle.
These are the foci of this review.

The paper is outlined as follows.
It starts by summarising the current exozodi literature (\S \ref{sec: synthesisOfLiterature}).
This includes a brief review of the observational constraints (\S \ref{subsec: summaryOfObsConstraints}),
and then a discussion of current models to explain warm and hot exozodis (\S \ref{subsec: warmExozodiModels}
and \S \ref{subsec: hotExozodiModels} respectively).
It then identifies 8 key questions that must be addressed to understand exozodis and their potential impact
on exo-Earth imaging missions (\S \ref{sec: theoInsightIntoKeyQuestions}).
For each question, current theoretical understanding is used to inform the answer to the question and to identify
outstanding issues such as where knowledge gaps might be.
Each question concludes with a brief {\it key finding} summarising the status of its resolution.
Conclusions are given in \S \ref{s:conclusion}.

\section{Synthesis of the literature}
\label{sec: synthesisOfLiterature}

\subsection{Summary of observational constraints}
\label{subsec: summaryOfObsConstraints}

This section briefly summarises what current observations of exozodis show.

\subsubsection{Detection wavelengths}

\begin{figure}[]
    \centering
    \includegraphics[width=8.5cm]{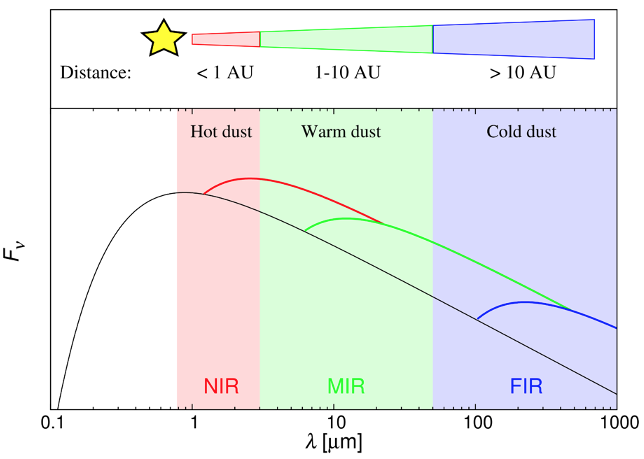}
    \caption{Illustration of the spatial location of dust in a planetary system and the wavelength at which
    its emission is seen in the spectral energy distribution (SED).
    (Top) Schematic showing the approximate locations of hot, warm and cold dust relative to a star.
    The exact locations are set by the stellar type; this plot corresponds to a Solar-type star.
    (Bottom) Corresponding SEDs of the dust populations (red, green, blue) on top of that of the star (black).
    The level of the dust emission is arbitrary, and typical hot and warm dust levels are significantly fainter
    than the star.
    Adapted from \citet{Kirchschlager2017} and reproduced by permission of Oxford University Press on behalf of
    the Royal Astronomical Society.}
    \label{fig:sf4}
\end{figure}

Warm exozodis are detected at mid-infrared wavelengths, usually in the \textit{N} band around ${10\um}$
(e.g., Fig.~\ref{fig:sf4}).
The brightest of these can be detected by photometry \citep[e.g.,][]{Kennedy2013}, while lower exozodi levels
can be reached with spectroscopy \citep[e.g.,][]{Chen2014}, and the faintest levels detected so far have
required use of nulling interferometry to disentangle the contribution of the stellar photosphere
(e.g., \citealt{Mennesson2014, Millan-Gabet2011, Ertel2018}).

Hot exozodis are commonly detected using optical long-baseline interferometry in the \textit{H}, \textit{K} and \textit{L} bands,
at wavelengths of order ${1\um}$ (e.g., \citealt{Absil2013, Ertel2014, Kirchschlager2020, Absil2021}; see Fig.~\ref{fig:sf4}).
Hot exozodis are not detected in mid-infrared observations, and no hot exozodi has been detected at wavelengths
of ${10\um}$ or longer \citep{Pearce2022Comets}.
This sets constraints on the properties of hot exozodi dust and indicates a lack of correlation between hot and warm exozodi
populations.

\subsubsection{Detection rates and brightness}
\label{sss:detectionRatesAndBrightness}

Surveys like HOSTS find that about 20 to ${30\%}$ of main-sequence stars have detectable warm exozodis \citep{Ertel2020}.
A similar fraction host hot exozodis, although hot and warm exozodis do not necessarily exist around the \textit{same} stars.
Of course this rate is just what can be seen above the detection threshold;
it might be expected that all stars have exozodi at some level.

The brightnesses of warm exozodis are often expressed relative to the Solar System's zodiacal cloud.
If 1 {\it zodi} is defined as the surface density of habitable-zone dust in the Solar System, then the surface densities
of detected warm exozodis are typically 10s to 1000s of zodis \citep{Ertel2020}.
Current detection limits are of the order of 10s of zodis, so only bright warm exozodis can be detected at present.

The brightness of a hot exozodi is often expressed differently, in terms of fractional excesses.
Typical hot-exozodi fluxes are around ${1\%}$ of the stellar flux in the $H$~or $K$~bands, which implies they are a
very bright feature of planetary systems.
This is close to the detection limit with modern instruments.
The brightness of at least one hot exozodi is known to vary by ${100\%}$ on a year-long timescale \citep{Ertel2016}.

\subsubsection{Dust location, size and temperature}

Most exozodis are detected via interferometry, allowing to constrain their location by spatially resolving the dust.

Warm exozodis are observed using nulling interferometry.
These observations are only sensitive to emission outside the inner working angle, which is typically of
order ${10\mas}$ from the star (e.g. \citealt{Ertel2018, Ertel2020, Pearce2022Comets}).
They are also only sensitive to emission inside the field of view, which is typically
around 0.5'' FWHM \citep{Mennesson2014, Ertel2018}.
This constrains typical warm exozodis to lie within a few au of stars.

The mid-infrared detections of warm exozodis imply that their spectral-energy distributions peak at around ${10\um}$.
This suggests that their temperatures are around ${300\K}$, placing them in the habitable zone,
which is consistent with the interferometric constraints.
These grains appear to be larger than several microns in size (e.g., \citealt{Lebreton2013}).

Conversely, hot exozodis are observed with interferometry at high spatial resolutions.
The emission profiles of hot exozodis seem to be very steep, because they are detected at near-infrared but not at
mid-infrared wavelengths.
This is confirmed by measurements of the steep spectral slope around ${3.5\um}$ for one system \citep{Kirchschlager2020}.
This steepness implies that the grains are very small and hot, typically sub-micron sizes with
temperatures of 1000 to ${2000\K}$ (e.g. \citealt{Lebreton2013, Kirchschlager2017}). 
These high temperatures suggest that the dust may be close to the star, which again agrees with the interferometric
constraints on location.
Some large grains may also be present in hot exozodis \citep{Stuber2023}, but small grains must dominate the size 
distribution (e.g. \citealt{Lebreton2013, Pearce2022Comets}).
The grains appear to be carbonaceous rather than silicate, because the lack of mid-infrared detections is
inconsistent with the silicate feature at ${10\um}$ \citep{Lebreton2013, Kirchschlager2017, Sezestre2019},
albeit requiring a composition which can survive at such high temperatures.
The emission appears to be mainly thermal, with minimal contributions from scattered light based on theoretical
modelling of grain emission and a lack of polarisation detections
(e.g. \citealt{vanLieshout2014, Rieke2016, Marshall2016, Kirchschlager2017}).
It is possible that scattered light contributes, particularly in the $H$~band \citep{Ertel2014}, but the
degree to which it does is still an open question.

\subsubsection{Correlations between exozodis and other observations}

Warm exozodis are strongly correlated with the presence of cold dust
which is found at 10s of au from the star (see Fig.~\ref{fig:sf4}).
\cite{Mennesson2014} found a statistical correlation between warm-exozodi detections and the presence of detected cold
dust, and more recently the HOSTS survey confirmed this trend \citep{Ertel2020}.
In the HOSTS sample, warm exozodis were detected around ${78^{+8}_{-18}\%}$ of stars that also had detected cold 
dust, compared with ${11^{+9}_{-3}\%}$ for stars without detected cold dust.

Conversely, there are no clear correlations between hot exozodis and either warm exozodis or cold dust (e.g. \citealt{Mennesson2014, 
Millan-Gabet2011, Ertel2014, Ertel2018, Ertel2020, Absil2021}).
There may be tentative trends; ${50\pm16\%}$ of HOSTS systems with a hot exozodi also have a warm exozodi,
whilst just ${20^{+6}_{-12}\%}$ of those without a hot exozodi also have a warm exozodi \citep{Ertel2020}.
There may also be a tentative correlation between hot dust and exocomet activity \citep{Rebollido2020}.
However, these trends are much less significant than the clear correlation between warm exozodis and cold dust.

\begin{figure}[]
    \centering
    \includegraphics[width=8.5cm]{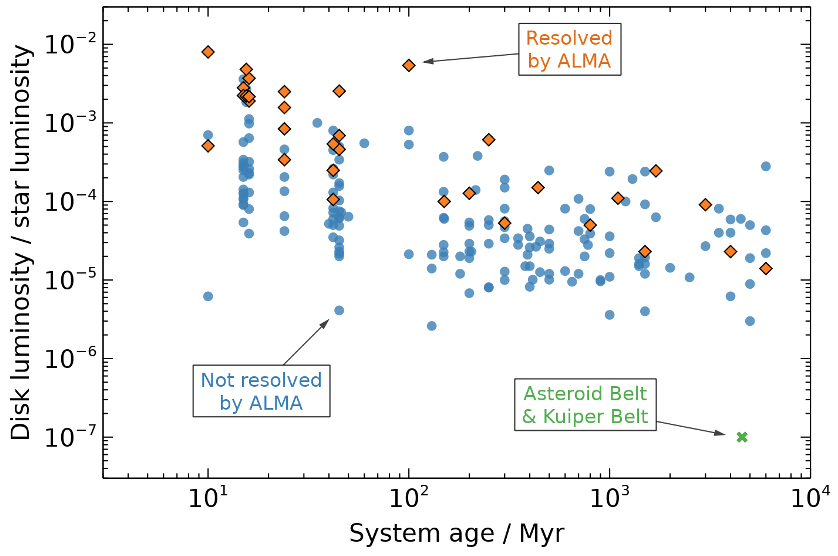}
    \caption{The evolution of cold debris disk luminosities \citep[adapted from][]{Pearce2024},
    showing the decline in disk brightness with age due to collisional erosion.
    Orange diamonds are ALMA-resolved disks, whilst blue points are from SEDs.
    The green cross shows the combined Asteroid Belt and Kuiper Belt, which are much less luminous than detected
    extrasolar debris disks.
    }
    \label{fig:sf5}
\end{figure}

There are also no clear correlations between either warm or hot exozodis and the star's age or spectral type, or with the presence of 
known planets in the system.
Both types of exozodi are found around A-type to K-type stars, with ages ranging from 10s of Myr to several Gyr
\citep{Kirchschlager2017, Ertel2020}.
This is in contrast to the clear age dependence found for colder debris \citep[see Fig.~\ref{fig:sf5}, e.g.,][]{Su2006},
although current searches for correlations are limited by the small number of stars surveyed for hot exozodis.

\subsection{Summary of models to explain warm exozodis}
\label{subsec: warmExozodiModels}

\begin{figure}[]
    \centering
    \includegraphics[width=8.5cm]{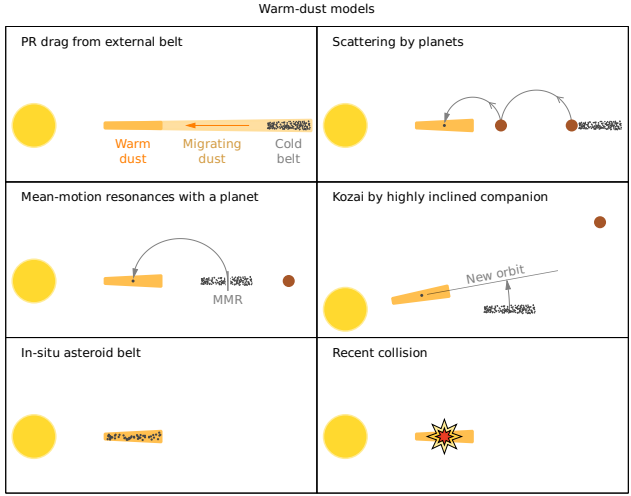}
    \caption{Cartoon summarising the models proposed to explain warm exozodis in the habitable zones
    of stars, as described in \S \ref{subsec: warmExozodiModels}.
    Each panel shows a different model.
    The large yellow circle is the star, the orange wedge the warm exozodi,
    dark grey points are planetesimals and brown circles are planets.
    Arrows denote migration pathways.
    }
    \label{fig:cartoonwarm}
\end{figure}

This section discusses models proposed to explain warm exozodi, as summarised in Fig.~\ref{fig:cartoonwarm}.

\subsubsection{Overview of modelling approaches}

Before considering the different models that have been proposed to explain warm exozodis, it is worth noting the different modelling 
approaches that have been employed to make predictions for the level of exozodi expected in different scenarios. The reason for the 
multiplicity of approaches is the range of different physical processes at play. Most notably there is a tension between the treatment 
of dynamical and collisional processes. The former can be followed accurately using N-body codes, and include gravitational forces 
acting on debris from the star and planets, as well as radiation forces. Collisional processes, on the other hand, need to be followed 
probabilistically, since individual collisions are rare, and when they occur multiple fragments are produced making it impossible to 
follow all components in a reasonable computational time.

Broadly speaking exozodi models can be classed into those that take N-body simulations as their basis (i.e., following the motion of 
individual bodies in the disk), and then approximate the effect of collisions (e.g., using collisional grooming or super-particle 
techniques, or look-up tables for collision outcomes; \citealt{Stark2009}, \citealt{Kral2015}, \citealt{Watt2024}), and those which 
take kinetic models as their basis (i.e., following the number of bodies in a certain size range and part of physical or orbital 
element parameter space; \citealt{Krivov2005}) in which case the dynamics may be approximated (e.g., by considering planet 
perturbations as advection terms). Both approaches are computationally intensive and so a third class of models is one which uses 
analytical calculations to follow the dynamical or collisional evolution (or a combination of both), being then faster to compute.

Perhaps frustratingly, there is no one approach which models a debris disk's structure much better than others. All models come with 
the caveat that they are applicable only for a certain set of assumptions, which usually means having to ignore physical processes 
which are assumed to be irrelevant. For example, collisional grooming models account for dust destruction but not creation in 
collisions, an issue that is overcome in kinetic models but at the expense of not following planetary perturbations accurately. This 
caveat is not presented to discourage the reader, simply to illustrate how this field necessitates the development of bespoke 
modelling approaches that take advantage of the dominant physics in a given situation, and the range of initial conditions that need 
to be explored.

A further caveat is that there remain uncertainties in the way some physical processes are modelled. For example, collisional outcomes 
are usually encapsulated within a dispersal threshold parameter $Q_{\rm D}^\star$, which while well studied in certain regimes 
(experimentally for cm-sized dust, and numerically for large planetesimals), has large uncertainties for smaller and intermediate 
sizes. Dust optical properties, which determine radiation forces and a disk's appearance in observations, also have uncertainties. 
Some physical processes have simply yet to be studied in detail; e.g., the possibility of gas being created in collisions and then 
affecting the dust evolution. This leaves open the possibility that the next generation of models, by including better prescriptions 
for the range of relevant physics, will predict new types of structures, or at least result in quantitatively different predictions 
for the dust levels.

\subsubsection{Collisions in an outer belt + P-R drag}
\label{subsec: collisionInOuterBeltAndPRDrag}

Given that there is a correlation between the presence of a warm exozodi and the presence of a cold, outer belt (analogous our Kuiper 
Belt), the simplest model for the origin of an exozodi is that it is a natural component of a system's outer belt. In the simplest 
model planetesimals in that belt maintain their orbits around the star until they collide with other planetesimals, at which point 
they fragment creating a collisional cascade of smaller fragments which also just orbit the star until they are destroyed in a 
collision. Such a cascade goes down to the smallest $\mu$m-sized dust which is removed on a dynamical timescale by radiation pressure. 
This model readily reproduces observations of exo-Kuiper belts, such as the halo of dust seen exterior to the belt caused by radiation 
pressure \citep{Strubbe2006}, and the manner in which the belts are seen to be fainter around older stars which can be inferred to be 
due to collisional erosion \citep[][see Fig.~\ref{fig:sf5}]{Wyatt2007}.

However, Poynting-Robertson (P-R) drag should also transport the small dust inwards (see Fig.~\ref{fig:cartoonwarm} top left).
\cite{Wyatt2005} modelled this process 
analytically, deriving a simple formula for the radial distribution of dust interior to the belt in the simplified scenario when 
collisions produce dust all of the same size. The resulting surface density profile depends on the ratio $\eta_0$ of P-R drag lifetime 
to collision time in the belt: this profile is flat for $\eta_0 \ll 1$ as dust migrates in without suffering a collision (like in the 
zodiacal cloud) and is strongly depleted inside the belt if $\eta_0 \gg 1$ as dust is destroyed before migrating very far. At the time 
this was used to motivate why it is possible to ignore P-R drag in detectable debris disks since these must be so dense that 
collisions dominate. However, with the advent of observational techniques like nulling interferometry able to detect faint levels of 
dust interior to the exo-Kuiper belts it was realized that the observed dust level is actually close to that predicted by this model 
\citep{Mennesson2014}.

\begin{figure}[]
    \centering
    \includegraphics[width=8.5cm]{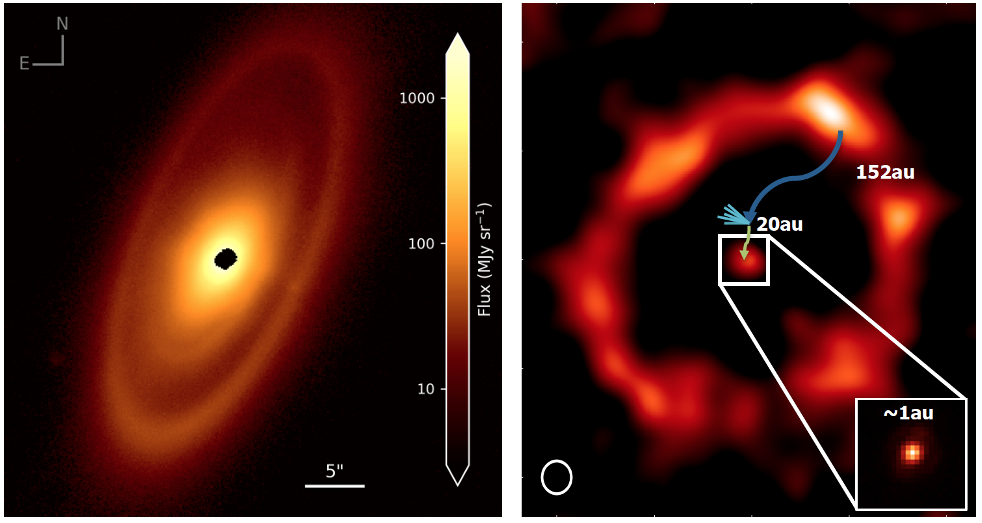}
    \caption{Images of debris disks in which cold outer belts are inferred to be supplying dust to the inner regions of the systems.
    (Left) JWST $25.5$\,$\mu$m image of the Fomalhaut debris disk \citep[adapted from][]{Sommer2025}.
    The outer belt is seen at $\sim 130$\,au with dust extending all the way in towards the star \citep{Gaspar2023} in a distribution
    that is consistent with inward transport due to P-R drag \citep{Sommer2025}.
    (Right) ALMA 870\,$\mu$m image of the $\eta$ Corvi debris disk \citep[adapted from][and reprinted with permission from Elsevier]{Wyatt2020}.
    The outer belt is seen at $\sim 150$\,au from which comets are inferred to be scattered in that sublimate at $\sim 20$\,au explaining
    CO gas detected there and the bright exozodi seen closer in at $\sim 1$\,au \citep{Marino2017}.
    }
    \label{fig:fomec}
\end{figure}

This meant that the simple models had to be updated. \cite{Kennedy2015_PRDrag} used the kinetic approach of \cite{vanLieshout2014} to 
model the full size and spatial distribution of dust due to planetesimal belt taking into account collisions and radiation forces, and 
used this to find an empirical correction factor to the \cite{Wyatt2005} model to get a more accurate surface density profile. This 
showed the parameter space within which parent exo-Kuiper belts must lie to populate detectable inner exozodis, showing that these 
colder outer belts could have evaded detection. This analytical model was again improved in \cite{Rigley2020} which showed that the 
full size and spatial distribution of the kinetic model could be approximated by each particle size having a spatial distribution 
characterized by its own $\eta_0$. The resulting analytical model now provides a reasonably accurate description of the size and 
spatial distribution of dust evolving only due to collisions and P-R drag, and has been tested by its ability to explain the spatially 
resolved dust emission seen interior to the outer belt in the Fomalhaut system \citep[see Fig.~\ref{fig:fomec} left;][]{Gaspar2023,Sommer2025}.
This model showed that the exozodi 
levels observed towards nearby stars in the HOSTS survey that also have outer belts \citep{Ertel2020} are close to the level predicted 
if they are replenished by P-R drag.

Nevertheless, the predicted dust level is slightly higher than expected and moreover there are systems like $\eta$ Corvi for which the 
dust level is far in excess of that expected (see Fig.~\ref{fig:fomec} right).

\subsubsection{Collisions in outer belt + P-R drag, dust also interacting with interior planets}

A natural extension of the simple model of \S \ref{subsec: collisionInOuterBeltAndPRDrag} is to consider a situation in which 
there are planets orbiting the star interior to the cold outer belt. Gravitational perturbations from such planets would disturb the 
inward migration of dust from the outer belt - some dust would be removed from the debris disk by being accreted by the planet, some 
would be ejected, while some would have its migration temporarily halted due to resonance trapping, causing an overdensity of dust 
outside the planet's orbit (see Fig.~\ref{fig:sf1}).

The processes of accretion and ejection have been well studied for planets on circular orbits using N-body simulations. 
\cite{Moro-Martin2003} showed how these processes can cause a drop in surface density interior to the planet's orbit. A more extensive 
parameter space exploration with more particles was later used by \cite{Bonsor2018} to derive empirical relations for the probability 
of different outcomes (i.e., ejection or accretion) for dust as it passes a planet. This showed how to predict the depth of gap carved 
by a planet, but a rule of thumb is that it requires a Saturn-mass planet or larger at 10s of au to deplete dust before reaching the 
inner region of a system. By the same argument, if an exozodi is seen and inferred to have its origin in P-R drag from an outer belt, 
then such massive planets cannot be present.

The process of resonance trapping received much attention because the Earth has such a resonant ring, and the special geometry of 
resonances means that this is clumpy \citep[][see Fig.~\ref{fig:sf1}]{Dermott1994}.
The different types of structures that might result were laid out in early 
papers like \cite{Ozernoy2000} and \cite{Kuchner2003} using N-body simulations, with \cite{Shannon2015} providing empirical relations 
for the probability of trapping in different resonances and the subsequent evolution and so their resulting structures.

While N-body simulations can provide an accurate representation of the dynamical structure of a resonant ring, the models of \S 
\ref{subsec: collisionInOuterBeltAndPRDrag} showed that consideration of collisional processes cannot be ignored. This led to the 
development of the collisional grooming models of \cite{Stark2008}. This showed how the previously inferred clumpy ring structures 
would still be expected, albeit at contrast levels to the background disk that require collisional models for an accurate prediction. 
\cite{Stark2011} showed the parameter space that maximises the level of resonant structure that can be expected, noting that most 
models focus on structures due to interactions in single planet systems, which may be weaker in multiplanet systems. Most recently 
\cite{Currie2023} presented a suite of simulations of resonant ring structures using this approach.

\subsubsection{Comet models: parent bodies from outer belt undergoing scattering by planets}

As well as migrating as dust, material from an outer planetesimal belt can also be transferred inwards in the form of comets
(see Fig.~\ref{fig:cartoonwarm} top right).
These comets would release dust in the inner regions, either via sublimation or by their disintegration, and so could thus sustain an 
exozodi. Comets appear to be the main mechanism sustaining the Solar System's zodi, and cometary models have advantages over P-R drag 
models in explaining hot exozodis (\S \ref{subsec: hotDustSupplyModels}).

A comet with apocentre in the outer regions and pericentre in the inner regions would heat up as it approaches the star. This would 
cause ices in the outer layers of the comet to sublimate, releasing gas and dust. For a comet comprising a large quantity of water 
ice, the rate of sublimation would significantly increase once the comet enters the habitable zone, and the corresponding increase in 
dust release would contribute to the exozodi. This comet-sublimation model can reproduce the exozodi levels in several systems 
\citep{Marboeuf2016}.

However, sublimation is not the dominant mass-loss mechanism for comets; much larger quantities of dust are released through cometary 
fragmentation, which accounts for the large majority of zodiacal dust in the Solar System \citep{Nesvorny2010}. In addition, the 
stochastic nature of cometary activity means that an exozodi produced by cometary fragmentation would be highly variable 
\citep{Rigley2022}. Cometary fragmentation could therefore be the main source of exozodical dust too.

If comets originating in an outer belt are to supply dust to the habitable zone, then some mechanism must continually drive comet 
pericentres down into the inner regions \citep[e.g.,][]{Levison1997}. Since exozodis are observed around systems with a broad range of 
ages, it would have to be possible for comets to be supplied at both early and late times. One way to achieve this is through various 
dynamical interactions with planets, with one possibility being inward scattering by a chain of planets. In this model, an outer 
planet located near the inner edge of a planetesimal belt scatters material onto eccentric orbits. Some of these orbits have 
pericentres interior to the planet, and this material may encounter another, inner planet. This inner planet then scatters the 
material again, driving some even further inwards and encountering more planets. In this way, a chain of planets can efficiently pass 
material inwards from an outer planetesimal belt to the habitable zone, where it could release dust to replenish an exozodi 
\citep{Bonsor2014, Raymond2014, Marino2018}, as inferred to be the case for $\eta$ Corvi \citep[see Fig.~\ref{fig:fomec} right;][]{Marino2017}.
Since the planetesimals have mass, this process typically causes the planetary orbits to 
diverge, with the outermost planet migrating outwards and the innermost migrating inwards. This means the outermost planet migrates 
into the disc, providing a fresh source of material and potentially allowing the process to continue throughout the star's lifetime 
\citep{Bonsor2014, Raymond2014}.

\subsubsection{Other dynamical scenarios that could transport outer bodies to the habitable zone}

There are several other mechanisms that could continually drive comets from an outer belt into the habitable zone, besides scattering 
by a chain of planets (see Fig.~\ref{fig:cartoonwarm} middle panels).
\citet{Faramaz2017} showed that a moderately eccentric planet located exterior to a planetesimal belt can drive 
material into the inner regions. In this model, material near internal mean-motion resonances gets excited to very high 
eccentricities, such that either their pericentres reach the inner regions of the system, or they pass close to the planet at which 
point they may get scattered inwards \citep[see also][]{Beust2024}. The process of resonantly exciting and then scattering material 
can take Gyr timescales, so this process could sustain an exozodi over very long times.

Other planetary interactions can also drive material from outer belts down into the habitable zone. A highly misaligned, eccentric 
companion exterior to a planetesimal belt can drive planetesimals to very high eccentricities through the Eccentric Kozai–Lidov 
Mechanism, bringing their pericentres into the inner regions \citep{Young2024}. Alternatively, an eccentric planet on a disc-crossing 
orbit, in a configuration similar to that of the tentative object Fomalhaut b, would drive debris pericentres down to the inner 
regions through secular interactions \citep{Pearce2021, Costa2024}. This process may take considerable time, so could sustain cometary 
inflow over long periods, although it may require a contrived dynamical setup to operate.

One scenario that seems unlikely to be responsible for exozodis is a system-wide dynamical instability, akin to the Late-Heavy 
Bombardment theory for the Solar System \citep{Booth2009}. In this model, multiple planets enter a configuration that is highly 
unstable, leading to a violent rearrangement of the planetary system. During this process, large quantities of planetesimals can be 
driven onto highly eccentric orbits that potentially bring them into the habitable zone. However, whilst such instabilities have 
previously been suggested as the source of exozodis, \citet{Bonsor2013} showed that the effect would be too short lived to be 
compatible with the majority of observations.

\subsubsection{In-situ asteroid belt}
\label{subsec: inSituAsteroidBelt}

Given that dust in the outer regions of planetary systems is thought to be replenished by collisions between planetesimals in Kuiper 
belt analogues, it is natural to expect that exozodiacal dust seen closer to the star might also be replenished from planetesimals 
that reside in an asteroid belt analogue (see Fig.~\ref{fig:sf1} and Fig.~\ref{fig:cartoonwarm} bottom left).
While any exo-asteroid belt would indeed create an exozodi, the dust level that it can 
sustain is restricted by the unavoidable collisional erosion of the belt's planetesimals (see \S \ref{subsec: 
collisionInOuterBeltAndPRDrag}).  This erosion works in such a way that once the belt is depleting (i.e., sufficient time has elapsed 
for the biggest bodies to be depleted in mutual collisions), there is a maximum mass of planetesimals that can remain that depends 
only on the age of the star and the radius of the belt. For a field star with an age of order ${1\gyr}$ this means that asteroid belts 
at ${1\au}$ should have eroded such that the dust they create is at levels far below the detection threshold by the current epoch 
\citep{Wyatt2007}. That is, any readily detectable warm exozodi cannot originate from an in-situ belt, but must be transiently 
replenished.

While some simplifications were needed to the analytical calculations that were used to propose a limit on the mass that can remain 
after a given duration of collisional processing, the existence of such a limit was found to be a robust feature of collisional models 
\citep{Heng2010}. It is, however, worth noting the strong radial dependence of the predictions, which mean that ${3-10\au}$ asteroid 
belts could nevertheless survive to replenish exozodis around Gyr-old stars. Thus despite the above challenge to in-situ belts, there 
is a possibility that they could produce cooler exozodis (e.g., \citealt{Su2013}).

\subsubsection{Recent collision}

The transience of the dust seen in warm exozodis (see \S \ref{subsec: inSituAsteroidBelt}), along with the fact that dust in 
debris disks is replenished in collisions, leads straight-forwardly to the possibility that the transient dust, at least in
some systems, is the product of a single recent collision (see Fig.~\ref{fig:cartoonwarm} bottom right),
rather than the collisional grinding of many planetesimals.
This is possible because the mass of dust required to be present for a detectable infrared signature only
corresponds to that of a modest-sized planetesimal (e.g., a 10s of 
km-sized planetesimal converted entirely into $\mu$m-sized dust at 1\,au would be detectable).
However, the viability of this explanation depends on some unknowns,
such as the size distribution of dust created in a collision, the lifetime of the resulting 
collision products, and the frequency of such collisions (which depends on the total mass of planetesimals which is a key unknown, 
e.g., \citealt{Krivov2021}).
These factors will determine the fraction of exozodis for which this is likely to be their origin, which is likely to some but
not all.

While it has long been expected that an outer planetesimal belt would be intrinsically clumpy due to individual collision events 
\citep{Wyatt2002}, it requires optimistic assumptions about the mass in the debris disk, and the efficiency of turning their mass into 
dust, for this to be a viable explanation for the clumpiness seen within belts \citep{Han2023}. Nevertheless, such collisions are 
inferred to explain morphological features in outer belts \citep{Stark2014, Jones2023, Rebollido2024} or time variability of the 
mid-IR emission \citep{Chen2024} and there is strong evidence for giant impacts (i.e., between planetary embryos) as the origin of 
warm dust in some systems. For example, the silica composition of the dust and recent CO production inferred for HD172555 
\citep{Lisse2009, Schneiderman2021} point to an event that would be comparable to that which formed the Moon \citep{Jackson2012}, and 
the carbonaceous composition inferred in other systems has been suggested to be linked to a giant impact \citep{Lisse2017}.

Such giant impacts are an inherent part of terrestrial planet formation models, taking place during a phase of instability lasting up 
to ${100\myr}$ after the protoplanetary disk has dispersed. Models for debris created in these events are complicated by the geometry 
– the debris is no longer axisymmetric, and the collision point is a special location through which all debris passes resulting in a 
high collision rate \citep{Jackson2014}. The high density can also lead to optical depth effects becoming important \citep{Su2019}, 
thus affecting the appearance of the debris cloud and its evolution. Collisions can also trigger avalanches of outward propagating 
spirals \citep{Grigorieva2007, Kral2013}.

While recent collisions are a plausible origin for exozodis around young stars, they may be a less viable explanation of dust
around older ($\gg 100$\,Myr) stars for which any population of planetary embryos would have already settled into a stable
planetary system and their asteroid belts would be depleted by collisional erosion.
However, the bright dust levels expected and asymmetry from the collision point might explain observations of warm exozodis 
to Gyr-old stars like $\eta$ Corvi and BD+20307 \citep{Defrere2015, Weinberger2011}.

Smaller scale impacts onto planets may also release dust into an exozodi. Indeed, impacts onto Mars have been suggested as an 
explanation for features in the zodiacal cloud \citep{Jorgensen2021}. While this interpretation has been disputed \citep{Pokorny2023}, 
planets are another potential source of debris in the habitable zone \citep[e.g.,][]{Wyatt2016}.

\subsection{Summary of models to explain hot exozodis}
\label{subsec: hotExozodiModels}

\begin{figure}[]
    \centering
    \includegraphics[width=8.5cm]{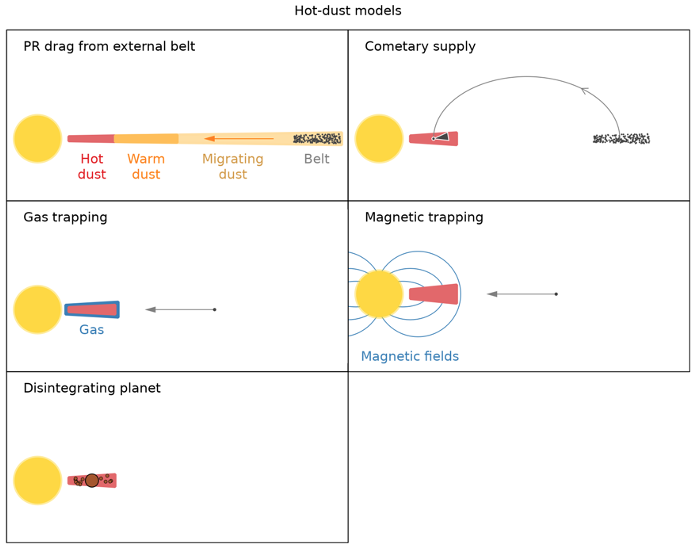}
    \caption{Cartoon summarising the models proposed to explain hot exozodis very close to
    stars, as described in \S \ref{subsec: hotExozodiModels}.
    Symbols have the same meanings as on Fig.~\ref{fig:cartoonwarm}.
    }
    \label{fig:cartoonhot}
\end{figure}

Unlike warm exozodis, hot exozodis are considerably harder to understand.
To date, no model has self-consistently explained hot exozodis for the full range of star types and ages about
which this phenomenon is observed.
This is because dust grains that are small enough, hot enough and close enough to the star to reproduce
hot-exozodi observations should be rapidly removed or destroyed by various dynamical processes.
Specifically the small, sub-blowout grains should be quickly blown away from the star due to radiation pressure,
and even if they were somehow protected from this, then they should be rapidly destroyed by sublimation or collisions.
This has led to the idea that some unknown mechanism replenishes and/or sustains hot dust.
In this section we briefly describe the main hot-dust models considered the literature (see Fig.~\ref{fig:cartoonhot}),
noting that a detailed review of hot-dust models can be found in \citet{Ertel2025}.

\subsubsection{Supply only models}
\label{subsec: hotDustSupplyModels}

One possibility for sustaining hot-dust populations is to continually replenish the dust as it is lost through the above processes. 
This has given rise to a class of theories known as `supply only' models.

One of the first suggestions for hot-dust replenishment was via a collisional cascade in an \textit{in-situ} planetesimal belt. 
However, this idea was quickly dismissed, because a belt so close to a star would collisionally deplete far too rapidly to explain 
observations \citep{Wyatt2007, Lebreton2013, Kral2017}.  Another attempt was the P-R drag pileup model (see Fig.~\ref{fig:cartoonhot} top left),
where P-R drag causes dust to migrate inwards to the hot-emission region from some distant source \citep{Krivov1998, Kobayashi2008, Kobayashi2009, vanLieshout2014, 
Sezestre2019}. However, this model also fails, for two main reasons. First, it produces too much mid-infrared emission; observed hot 
exozodis have significant near-infrared emission yet no detected mid-infrared emission \citep{Pearce2022Comets}, but P-R dust would 
produce far too much mid-infrared emission as it migrated through the warm-emission region. Second, P-R drag alone cannot sustain a 
population of hot grains that are smaller than the blowout size, but sub-blowout grains are inferred for many hot-dust stars 
\citep{Kirchschlager2017}.

A potentially more promising model is cometary supply (see Fig.~\ref{fig:cartoonhot} top right),
where dust is directly deposited in the hot-emission region by star-grazing 
comets \citep{Bonsor2014, Raymond2014, Marboeuf2016, Faramaz2017, Sezestre2019, Pearce2022Comets}. This has the advantage that it 
produces considerably less mid-infrared emission than the P-R drag model, and stochastic cometary infall would offer a natural 
explanation for the near-infrared variability seen in at least one system \citep{Ertel2014, Ertel2016}. However, comets alone do not 
seem capable of reproducing hot-exozodi observations unless the cometary inflow rate is unphysically high \citep{Pearce2022Comets}.

Another possibility is that a disintegrating inner planet could continually release hot dust
\citep[see Fig.~\ref{fig:cartoonhot} bottom left;][]{Lebreton2013}, although there are 
no known correlations between hot exozodis and detected planets \citep{Ertel2014}, and it is potentially difficult to argue that all 
of the ${20\%}$ of main-sequence stars with hot exozodis currently host disintegrating inner planets.

In summary, none of the current supply-only models seem able to reproduce all hot-exozodi observations. It appears that simply getting 
dust close to a star is not enough to produce a hot exozodi, because this dust should get rapidly removed or destroyed. This has led 
to an alternative class of models, described in \S \ref{ss:trapping}, which speculate that some additional mechanism prolongs the 
time that hot dust can reside near stars.

\subsubsection{Trapping models}
\label{ss:trapping}

The second class of hot-dust models are `trapping models'. These hypothesise that, once grains reach the hot-emission region, they 
encounter some physical mechanism that protects them from being removed or destroyed.

One possible trapping mechanism is magnetic trapping (see Fig.~\ref{fig:cartoonhot} middle right).
In this model, grains become charged and then trapped in stellar magnetic fields 
\citep{Czechowski2010, Su2013, Rieke2016, Stamm2019}. However, current magnetic-trapping models struggle because it is unclear whether 
they can significantly extend grain lifetimes beyond the sublimation timescale. There are also no significant correlations between 
hot-dust detections and magnetic-field strength or stellar-rotation rate, which would be expected in this model \citep{Kral2017, 
Kimura2020}. Further investigations into magnetic trapping are currently ongoing (Peronne et al., in prep.).
A second trapping mechanism is gas trapping (see Fig.~\ref{fig:cartoonhot} middle left),
where gas released by sublimating dust can trap incoming grains just outside the sublimation radius 
\citep{Lebreton2013, Pearce2020}. This model can reproduce hot-exozodi observations for Sun-like stars, but fails for A-type stars 
because it struggles to trap sub-blowout grains. A third trapping mechanism involving the Differential Doppler Effect (DDE) also 
appears unable to reproduce observations \citep{Sezestre2019}.

In summary, current trapping models also fail to fully reproduce observations of hot exozodis.
This is generally due to the difficulty in devising a mechanism that both holds grains against radiation pressure
and protects them from sublimation and collisions.
Given these challenges, it is possible that current models omit some key physics which allows dust to survive for long times, or even that 
near-infrared emission does not actually arise from hot dust, possibilities which are discussed in \citet{Ertel2025}.
For now, no model has satisfactorily explained hot exozodis, and the origin and nature of this phenomenon remains a mystery.

\subsubsection{Hot-dust sources}

Regardless of the nature of hot exozodis, it seems unlikely that hot dust originates \textit{in situ}. Instead, the general consensus 
is that this material originates elsewhere in the planetary system, and gets transported inwards towards the star. The most commonly 
considered origins are planetesimal belts further out in the system, analogous to the Asteroid Belt and Kuiper Belt in our Solar 
System. Material would leave these reservoirs in the form of dust, boulders or larger planetesimals, and travel inwards under the 
action of drag forces (e.g., P-R drag, stellar wind drag or Yarkovsky forces) or planetary interactions.

However, this hypothesis faces a significant problem; there are no significant correlations between detected near-infrared excesses 
and mid- or far-infrared excesses, which would be indicative of massive outer planetesimal belts. It is possible that some hot-dust 
systems host planetesimal belts that lie below detection limits, which are nonetheless still massive enough to sustain hot exozodis. 
It is also possible that the material originates even further out, in structures analogous to the Oort cloud, and is supplied to the 
innermost regions in the form of long-period comets. The mass required in the source reservoir could also be reduced if an efficient 
hot-dust trapping mechanism operates, which would extend the lifetime of hot dust near stars and hence reduce the required inflow 
rate.

The other main possibility for the hot-dust source is disintegrating planets. However, as already noted above, it seems unlikely that 
the ${\sim20\%}$ of main-sequence stars with near-infrared excesses all host such planets. Distant planetesimal reservoirs therefore 
appear to be the most likely source, even if they must be too faint to detect with current instruments.

\subsection{White dwarf debris disks}
\label{ss:wd}

While this review is focussed on dust found in proximity to main sequence stars and the theoretical modelling of such exozodiacal 
dust, it is worth acknowledging the existence of dust in close proximity to white dwarfs and commenting briefly on the similarity (and 
differences) in the studies of these two types of object.

\begin{figure}[]
    \centering
    \includegraphics[width=8.5cm]{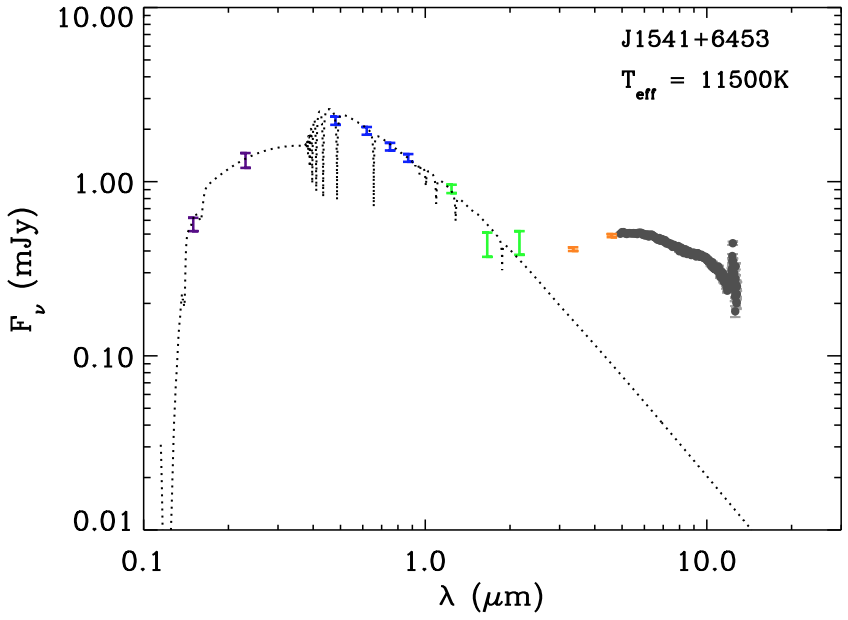}
    \caption{Spectral energy distribution of the white dwarf J1541+6453 \citep{Farihi2025}.
    Excess emission above the stellar photosphere is detected both photometrically with WISE (orange points) and
    spectroscopically with JWST (black points) indicating that the star hosts a $\sim 980$\,K dusty debris disk.
    }
    \label{fig:wd}
\end{figure}

White dwarfs are post-main sequence stars and so the descendants of main sequence stars. Observationally their debris disks are found 
using similar methods, albeit that such stars are fainter preventing use of interferometric methods. Thus this debris is usually found 
photometrically and is seen to have dust temperatures of $\sim 1000\K$ (see Fig.~\ref{fig:wd}),
although some white dwarfs have cooler dust in their habitable 
zones $\sim 300\K$ \citep[e.g.,][]{Jura2003}. While only detected towards a few percent of white dwarfs, such disks are likely much 
more ubiquitous given the metal pollution seen in $\sim 30$\% of white dwarf atmospheres.

The theoretical models that are used to interpret these observations share many similarities with the exozodi models discussed in
\S \ref{subsec: warmExozodiModels} and \S \ref{subsec: hotExozodiModels}.
That is, it is inferred that these stars have a planetary system at larger distances from the star that contains 
planetesimals and other debris, some of which ends up on orbits that bring them close to the stars \citep{Farihi2016}. This material 
replenishes the debris disk that is seen photometrically, before accreting onto the star where it can be detected in the star's 
atmosphere. The dynamical processes which have been considered to transport material in towards the star are the same as those 
considered to replenish exozodis, such as scattering by planets and binary star interactions \citep[e.g.,][]{Bonsor2015}. The same 
physical processes also affect the debris, such as collisional erosion and radiation forces acting on the dust 
\citep[e.g.,][]{Kenyon2017}.

There are some differences, however. For example, the high density of a white dwarf facilitates the break-up of planetesimals into 
smaller fragments by tidal destruction \citep[e.g.,][]{Malamud2020}. This can only occur when planetesimals reach very close to a main 
sequence star's surface, whereas for a white dwarf there is a large volume within which tidally disrupted debris can orbit, and tidal 
forces are often inferred to be the mechanism forming the debris disk \citep[e.g.,][]{Brouwers2022}. While the detailed physics of a 
white dwarf's debris disk remains unknown, the star's low luminosity prevents removal by radiation pressure, meaning that viscous 
processes, P-R drag or gas drag likely operate to remove dust \citep[e.g.,][]{Rafikov2011}, as well as sublimation which replenishes a 
gas disk that is also seen for some white dwarfs.

It is worth noting that between the main sequence and white dwarf phases, a star will have engulfed both any pre-existing exozodi and 
what will eventually become the white dwarf debris regions. That intermediate phase also includes significant loss in stellar mass 
which will lead to expansion and increased instability of planetary orbits \citep[e.g.,][]{Debes2002, Veras2013}, as well as a 
short-lived phase of high luminosity which would have destroyed small planetesimals. Thus there are many reasons why, despite their 
similarities, exozodis and white dwarf debris may not be the exact same phenomenon seen at different stages. Nevertheless, those 
similarities mean that there is a lot that studies of these objects can learn from each other.

\section{Theoretical insights into key questions}
\label{sec: theoInsightIntoKeyQuestions}

The aim of this section is to use the models of \S \ref{sec: synthesisOfLiterature} to inform on several key questions
about exozodis.
Specifically the goal here is to describe the contribution of our current understanding of the theoretical models to 
answering those questions.

\subsection{What is the size and spatial distribution of dust in exozodis?}
\label{ss:sizespatial}

As described in \mbox{\S \ref{sss:detectionRatesAndBrightness}}, in the absence of hard observational constraints on the 
structure of exozodis, warm exozodis are often parameterized in units of \textit{zodi}, where 1 zodi corresponds to the Solar System’s 
zodiacal cloud (see Fig.~\ref{fig:sf1}).
This means that the surface density is assumed to be fairly flat (i.e., constant with distance), with a 1 zodi disk 
having a geometrical optical depth ${\tau = 0.7 \times 10^{-7}}$ at 1\,au. To use such a model to make predictions for the thermal 
emission or scattered light from the disk would require knowledge of the size distribution and optical properties of the particles. 
However, for the simple models black body emission is assumed; e.g., see \cite{Kennedy2015_LBTI} for a description of the model used 
to interpret the HOSTS survey.

\begin{figure*}[]
    \centering
    \includegraphics[width=18cm]{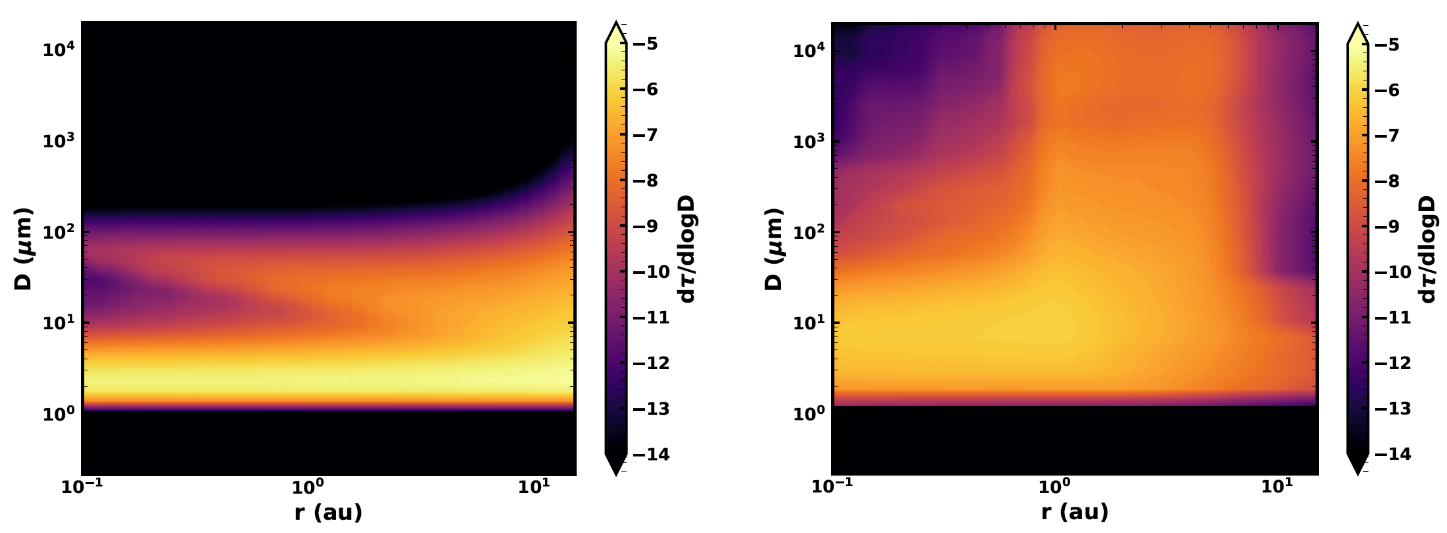}
    \caption{Size and spatial distribution of dust in two exozodi models, both of which have the same optical depth at 1\,au
    of 15 times the level in the zodiacal cloud (i.e., $\tau=1.1 \times 10^{-6}$), but different physical origins.
    {\it Left:} P-R drag model of dust dragged in from a $3.5 \times 10^{-5}M_\oplus$ belt at 30\,au \citep{Rigley2020}.
    {\it Right:} Comet model for the Solar system’s zodiacal cloud at an epoch of high dust levels following the scattering
    in of a long-lived massive comet \citep{Rigley2022}.}
    \label{fig: rigleyWyatt2022Fig}
\end{figure*}

While such simple models provide an efficient way to parameterize a warm exozodi, the true disk is likely to be a lot more 
complicated. For example, sticking with the example of the Solar system and our understanding from \S \ref{subsec: 
collisionInOuterBeltAndPRDrag}, we know that if the mass input rate to the zodiacal cloud was increased by a factor of 10 then this 
would change the competition between collisions and P-R drag. This would cause the inner regions to be more depleted so the surface 
density distribution may no longer be flat. It would also change the size distribution so that the total cross-sectional area of dust 
in the disk would not scale linearly with mass input rate. That is, what the simple models call a 10 zodi disk is not the same as a 
zodiacal cloud with a 10 times larger mass-input rate, the discrepancy becoming much greater as input rate increases. Added to this is 
the fact that the radial structure of an exozodi will depend on the architecture of its planetary system, which could be very 
different to the Solar system.

Theoretical models can be used to overcome these challenges to some extent, since they make quite detailed predictions for the size 
and spatial distribution of material in an exozodi, albeit that a prediction for a given system is inevitably limited by factors many 
of which will be poorly constrained (like the planetary system architecture). Nevertheless it is important to be aware that the origin 
of the exozodi has a strong effect on the size and spatial distribution, as illustrated in Fig. \ref{fig: rigleyWyatt2022Fig}. This 
shows two models, both of which are chosen to have the same surface density at 1\,au of 15 zodi, but in one of the models the dust is 
dragged in by P-R drag from an outer belt, whereas in the other the exozodi is replenished by comet scattering. This results in a 
different radial profile for the surface density in the two models, as well as a different conversion from surface density to thermal 
emission or scattered light brightness (because of the different size distributions), a point which is considered further in \S 
\ref{subsec: howDustSizeCompAffectExozodiObs}.

While the above discussion is focussed on the interpretation of warm exozodis, the general points it raises apply equally to hot 
exozodis. These can be summarised in the first key finding.

{\bf Key Finding 1:} The size and spatial distribution of dust in an exozodi is strongly dependent on its origin.

\subsection{What is the dominant exozodi delivery mechanism?}

The dominant origin of dust in exozodis, whether warm or hot, remains unsolved. However, it is possible to rule out scenarios in some 
cases, and an exozodi's morphology or composition can be found to be indicative of particular scenarios in others. The easiest way to 
rule out origin models of warm exozodis is based on the dust level, and so this gives the best guideline on which models might be in 
play in a given system. Below we will quantify this in terms of the fractional excess seen at 12\,$\mu$m ($R_{12}$), e.g., as defined 
in the exozodi luminosity function of Kennedy \& Wyatt (2013), but note that this quantification should only be regarded as 
approximate.

For example, an \textit{in-situ} asteroid belt is ruled out for all but the faintest warm exozodi dust levels (which would be below 
current detection thresholds, $R_{12}<0.01$), or youngest systems $<100$\,Myr. If a cold outer belt is present then P-R drag may be 
responsible, and indeed is a plausible origin of several known warm exozodis. However, this could only work for dust levels up to some 
maximum; this maximum depends on the outer belt's properties, but it is around the current detection threshold ($R_{12} \approx 
0.03$). A warm exozodi that is fed by P-R drag has a characteristic radial profile, generally being quite flat far interior to the 
outer belt, so this can be used to assess this possibility \citep[e.g.,][]{Sommer2025}.
However, this profile can be altered by the presence of intervening 
planets. It should also be noted that detectable exozodis can be fed by outer belts that have evaded detection.

For brighter warm exozodis (i.e., those readily detectable with $R_{12}>0.1$) alternative sources of dust are required. Exocomets are 
a plausible source for bright exozodis, but this requires an outer belt to act as a source for the exocomets and a planetary system 
architecture that enables efficient inward scattering. This is likely to limit the maximum exozodi level that may be produced in this 
way to $R_{12} < 0.3$. The radial profile of such an exozodi would depend on the planetary system architecture and the mechanism 
disrupting the exocomets, with the latter likely leading to exozodis concentrated in the inner regions where exocomets would be 
subject to thermal stresses. Gas released from exocomet sublimation may also be present, as inferred in one system
\citep[see Fig.~\ref{fig:fomec} right;][]{Marino2017}.

For the brightest exozodis ($R_{12}>0.3$) the favoured interpretation would be a recent collision between planetary embryos. The fact 
that the brightest exozodis in terms of fractional excess are found predominantly around stars with ages $<100$\,Myr 
\citep{Kennedy2013}, which is when planet formation models predict giant impacts to occur, favours this origin for such systems. It 
should not be forgotten that in-situ asteroid belts may be a plausible origin of bright warm exozodis for young systems. However, 
there is further compositional evidence in some systems that favours a giant impact origin, from the presence of silica dust created 
in a hypervelocity impact and short-lived CO thought to be stripped from the planetary atmosphere.

Note that the above division does not mean that faint warm exozodis, which are (likely) the most common, must form from in-situ 
asteroid belts. Rather such exozodis could be explained by any of the proposed origins. Hence the dominant origin is not yet known. 
This leads to the second key finding.

\textbf{Key Finding 2:} The origin of warm exozodis can in principle be determined for the brightest examples, however for fainter 
(but still problematic) dust levels all models are possible. Detailed characterisation of these (e.g., by measuring the radial profile 
or dust composition or size distribution) can be used to distinguish between models.

\subsection{How do exozodis pinpoint planets?}

Much of the literature has focussed on the negative impact of exozodis on future exo-Earth imaging. However, exozodis could also 
facilitate exoplanet detection and characterisation, because those planets may create detectable features in exozodi distributions. 
This is similar to how the presence and evolution of unseen planets can be inferred from observed features in cold debris discs at 10s 
or 100s \mbox{of au} \citep[e.g.,][]{Wyatt1999, Pearce2022ISPY}.
On the other hand, planet-induced features in exozodis could \textit{themselves} be mistaken for planets,
leading to their mischaracterisation \citep{Savransky2009}.

\begin{figure}[]
    \centering
    \includegraphics[width=0.4\textwidth]{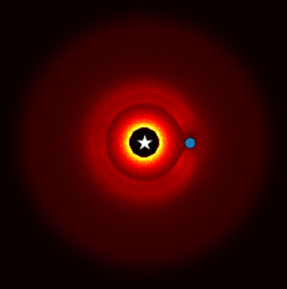}
    \caption{Simulation of the effect of an Earth-like planet at ${1\au}$ on an exozodi \citep[reproduced from][]{Currie2023}.
    The exozodi features a horseshoe-like structure, which arises from mean-motion resonances and is symmetrical
    about the planet (blue circle).
    The simulation shows scattered light from the exozodi, where the horseshoe represents a brightness
    increase by a factor of $\sim2$.
    }
    \label{fig: planetEffectOfExozodiCurrie2023}
\end{figure}

Several works have simulated the dynamical effect of habitable-zone exoplanets on exozodis (e.g. \citealt{Stark2008, Currie2023}). 
Those simulated exozodis often have clumps, induced by mean-motion resonances with planets.
These resonant structures would be strongly indicative of the planet's position, and would orbit the star with the planet.
Another common feature is a gap in the radial distribution at a planet's location, which forms as the planet
destabilises and scatters nearby material.
The magnitudes of both clumps and gaps could be exaggerated if planets migrate, or if grains move inwards through P-R drag.
An example simulation of a planet-exozodi interaction is shown in Figure \ref{fig: planetEffectOfExozodiCurrie2023},
which is a more extreme version of the Earth's resonant ring structure that is present in the zodiacal
cloud Fig.~\ref{fig:sf1}.

Planet-induced features in exozodis present both a problem and an opportunity for exoplanet imaging. On the one hand, detecting exozodi 
features like those in Figure~\ref{fig: planetEffectOfExozodiCurrie2023} would help to pinpoint the planet. The strength and shape of 
these features depend on planet's mass and orbit, so such observations would also help to characterise the planet. However, the 
possibility of mistaking clumps for planets also poses a problem, particularly if a clump is brighter than the associated planet. 
These clumps could potentially be distinguished from planets by taking spectral data, or by improved spatial resolution (i.e., since a 
planet would appear more point-like), but nonetheless may increase the number of false-positive exoplanet detections 
\citep{Savransky2009}.

\textbf{Key Finding 3:} Structures in exozodis can be used to pinpoint planets, but could also be confused for planets.

\subsection{How do exozodis affect a planet's physical properties and habitability?}

Exozodis may not only affect the detectability of planets, but also their physical properties. A planet embedded in an exozodi would 
accrete material, either in the form of dust, or as larger planetesimals like comets. Depending on the exozodi environment and the 
planet properties, the accreted material could significantly change the planet's properties and habitability.

One consequence of this accretion is the delivery of volatiles to habitable-zone planets. Earth's water may have been delivered in 
impacts of comets or outer main belt asteroids \citep{Sarafian2014, OBrien2018}, and this water is a major factor in Earth's 
habitability. Since exocomets are a likely source of exozodis \citep{Rigley2022}, the presence of a bright exozodi could imply high 
levels of exocometary activity and hence water delivery onto habitable-zone planets. The bulk composition of dust grains in an exozodi 
could differ from that of the larger planetesimals, e.g., with the latter better able to lock up volatiles. Thus dust accretion could 
deliver a slightly different set of substances and/or deliver it to a different part of the planet (e.g., the dust may disintegrate 
before reaching the surface). In general, the specifics of how accreted material changes a planet's composition would depend on the 
size and composition of impactors.

Another consequence of this accretion is to alter the planet's atmosphere. Bombardment can erode the primordial atmosphere that a 
planet was born with, while at the same time any accreted volatiles can be released to form a secondary atmosphere 
\citep{Wyatt2020Atmospheres}. Similarly, planetary atmospheres can be significantly altered through accretion of gas from the 
circumstellar environment. Such a gas disk could be replenished through ongoing planetesimal collisions in a cold outer debris belt, 
and this gas could viscously spread inwards where it could potentially accrete onto habitable zone planets. Hence such gas accretion 
could persist well beyond the protoplanetary disc phase \citep{Kral2020}. Since exozodis are strongly correlated with the presence of 
cold outer dust (and may be replenished by them), and many cold discs have detected gas indicating these are made of volatile-rich 
bodies, systems with exozodis may also be signatures of second-generation gas (either now or in the past). This gas could be that 
created in an outer belt as discussed above, but could also be created within the exozodi itself, and could alter (or have altered) 
planetary atmospheres.

Exozodis may therefore present both problems and opportunities in the search for habitable Earth-like planets. On the one hand, 
exozodis may make detection of such planets difficult. On the other hand, exozodis may signify cometary activity in the habitable 
zone, which would bring volatiles like water to these planets and hence increase their habitability. This might make planets in 
exozodi systems compelling candidates in the search for life, despite the additional difficulties in detecting planets embedded in 
exozodis. In any case, information about cold belts or outer planets can help to constrain a planetary system's dynamical history, 
also potentially letting us estimate the bombardment history of any habitable-zone planets, which is an important factor in assessing 
habitability in the system.

\textbf{Key Finding 4:} Exozodis define the environment within which habitable-zone planets reside, and could affect their physical 
properties and habitability.

\subsection{How do dust size and composition affect exozodi observable properties?}
\label{subsec: howDustSizeCompAffectExozodiObs}

\begin{figure*}[]
    \centering
      \includegraphics[width=0.7\textwidth]{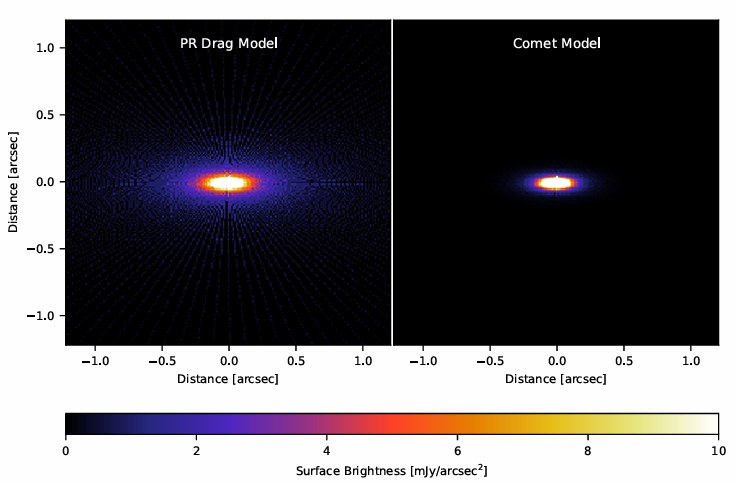}
      \includegraphics[width=0.7\textwidth]{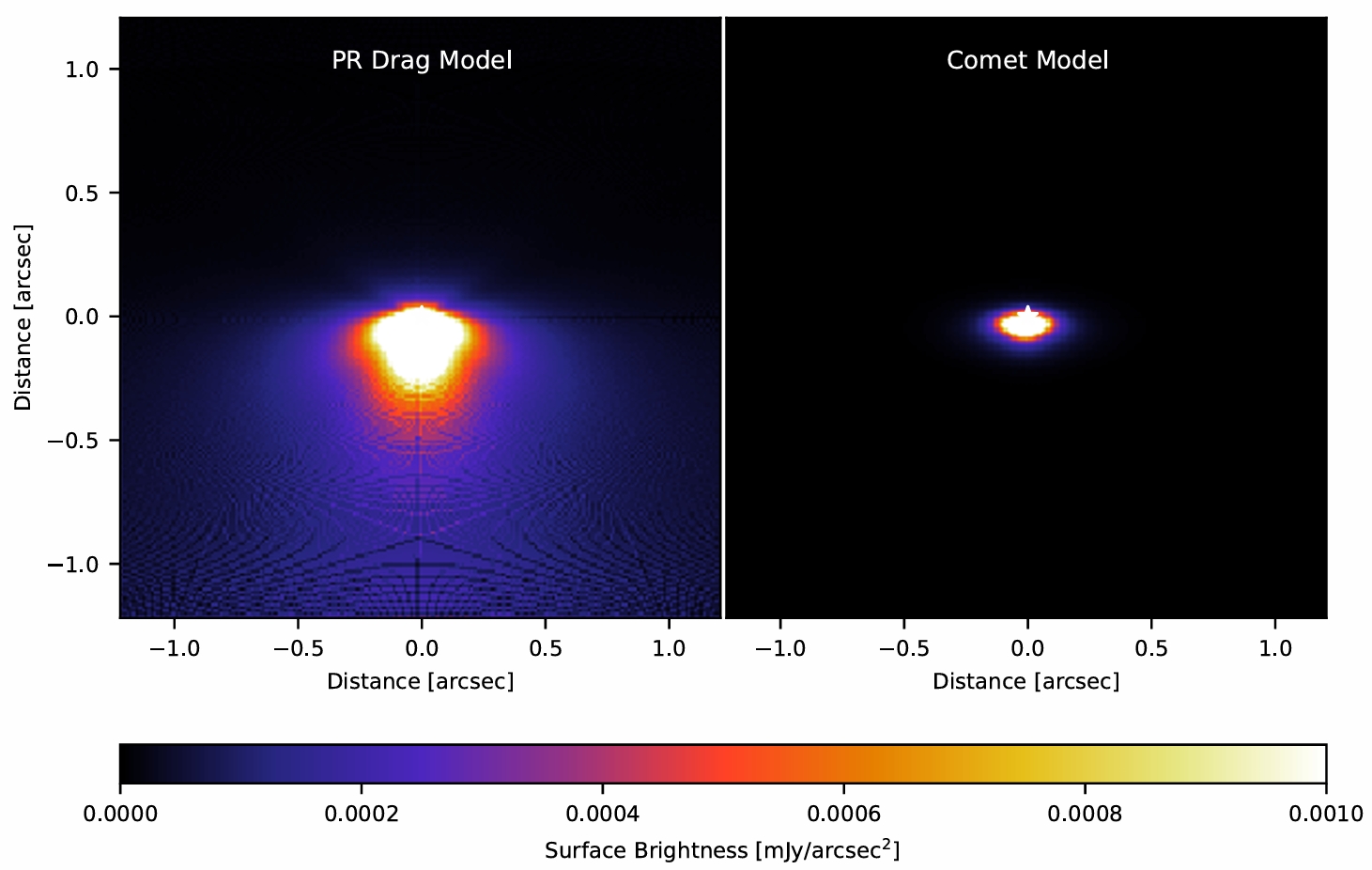}
    \caption{Simulated images of the model exozodis of Fig.~\ref{fig: rigleyWyatt2022Fig} placed around a Sun-like star
    at 10\,pc, when observed at $70^\circ$ inclination to the line-of-sight, assuming dust with an astronomical
    silicate composition and optical properties calculated using Mie theory.
    {\it Top:} Thermal emission images at $\lambda=12$\,$\mu$m wavelength for the P-R drag model
    \citep[left,][]{Rigley2020} and the comet model \citep[right,][]{Rigley2022}.
    {\it Bottom:} Scattered light images for the same models assuming $\lambda=1$\,$\mu$m.}
    \label{fig:Model_Images}
\end{figure*}

The observable properties of an exozodi depend not only on the dust sizes and spatial distribution, but also on the dust's optical 
properties.
These optical properties are determined by the dust composition, and are characterised by dust grains' absorption and
scattering coefficients.
These coefficients determine the amount of light at different wavelengths that is absorbed or scattered.
The light which is absorbed goes into heating the dust grains with that energy ultimately getting re-radiated in the infrared,
while some of that which is scattered is sent in our direction.
Literature works often assume Mie theory with compact spherical grains to determine the absorption coefficients
and so thermal emission properties of the dust \citep{bohren-huffman-1983}.
Consideration of a disk's scattered light properties may also use Mie theory, but more sophisticated approaches
such as Distribution of Hollow Spheres (DHS) or Discrete Dipole Approximation (DDA) that take into account
different shapes of particles might lead to more realistic results.

\begin{figure*}[]
    \centering
      \includegraphics[width=0.47\textwidth]{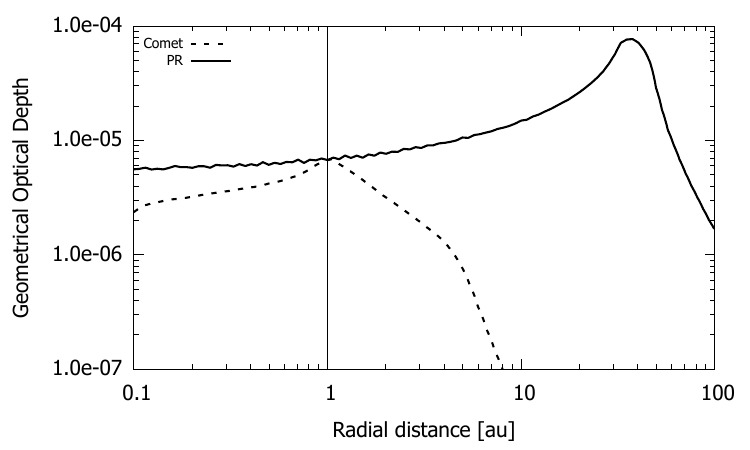}
      \includegraphics[width=0.47\textwidth]{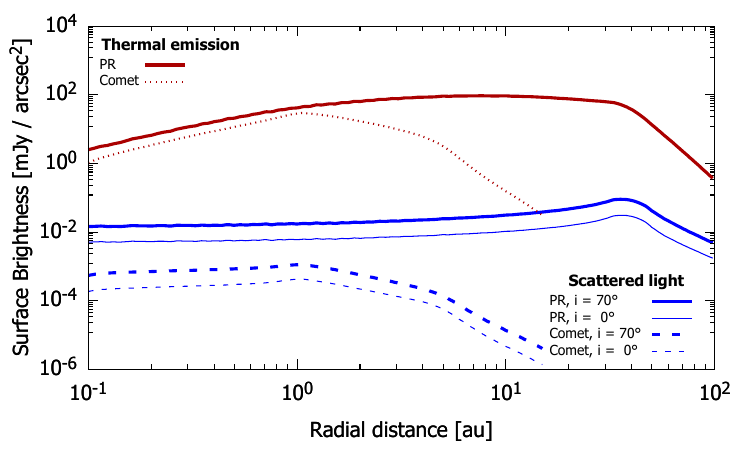}
    \caption{Radial profiles for the model exozodis of Figs.~\ref{fig: rigleyWyatt2022Fig} and \ref{fig:Model_Images}.
    {\it Left:} Geometrical optical depth as function of radius for the P-R drag model 
    \citep[solid line,][]{Rigley2020} and the comet model \citep[dashed line,][]{Rigley2022}.
    {\it Right:} Surface Brightness profiles for the P-R drag (solid lines) and the comet models (dashed lines)
    for both the thermal emission at $\lambda=12$\,$\mu$m (red) and scattered light at $\lambda=1$\,$\mu$m (blue).
    For scattered light, thick lines are the azimuthally averaged brightness of a disk viewed at $70^\circ$ inclination;
    thin lines are for $0^\circ$ inclination.}
    \label{fig:optical_depth}
\end{figure*}

It is worth acknowledging upfront that the dynamical models are not completely detached from the optical properties.
This is because the radiation force exerted on a dust particle (which is characterised by the $\beta$ parameter) is
determined by its absorption and scattering coefficients, which thus determine the size of particle corresponding to
a given $\beta$.
Knowledge of dust optical properties is thus crucial for connecting dynamical processes
(which are set by $\beta$) with collisional processes (which are set by particle size).
Ideally the dynamical models would take dust composition as input to allow a self-consistent consideration of the 
dynamics and the resulting observable properties of the disk.
However, for all but the smallest particles close to the blow-out limit there is a good analytical approximation
for the relation between $\beta$ and particle size which is often used \citep[e.g.,][]{Burns1979}.
While this allows the dynamics to be decoupled from the observable properties, this could introduce 
inaccuracies when considering observations that are dominated by those smallest particles.

Fig.~\ref{fig:Model_Images} shows an example of a conversion from dynamical models to observables, i.e., thermal emission and 
scattered light images.
This used the P-R drag and comet models \citep{Rigley2020, Rigley2022} from Fig.~\ref{fig: rigleyWyatt2022Fig}, which is
replotted in the left panel of Fig.~\ref{fig:optical_depth}, summed over particle size to show the optical depth as function of radius.
Both of these models were fixed to the same optical depth at a radial distance of 1~au.
This was then converted to surface brightness maps using a radiative transfer model for an assumed viewing geometry at wavelengths
of 1~$\mu$m (scattered light) and 12~$\mu$m (thermal emission), with the resulting images shown in
Fig.~\ref{fig:Model_Images}.
The same information is summarised on the right panel of Fig.~\ref{fig:optical_depth} for ease of comparison.
Both models assume Mie theory and a distance of 10~pc to the observer
\citep[for details on the radiative transfer model applied see][]{pawellek-et-al-2023}.
These figures show that while in thermal emission the different models lead to similar surface brightnesses within
1\,au, the values differ by an order of magnitude when considering scattered light.
This is a direct consequence of the different size distribution in the two models.
It illustrates the importance of understanding exozodi origins, for example when translating exozodi levels
detected in thermal emission by LBTI to the levels of scattered light;
the latter could act as confusion to a mission aimed at detecting exo-Earths at shorter wavelengths, like the 
Habitable Worlds Observatory.

While dust composition was kept the same in the models presented in Fig.~\ref{fig:optical_depth}, this would also be an important 
factor in determining an exozodi's observable properties. For example, the temperature of dust at the same distance from the star has 
a strong dependence on particle size due to the way the grains' absorption efficiency affects their temperatures, with smaller grains 
being hotter than larger ones \citep[e.g.,][]{pawellek-krivov-2015}. Different materials have different absorption efficiencies 
leading to potentially significant differences in thermal emission surface brightness. Similarly the scattering properties of a grain 
are determined by the material's optical constants and the shape of the particles. Here, also the porosity of the dust material plays 
a significant role \citep[e.g.,][]{arnold-et-al-2019, pawellek-et-al-2023}. Dust size, shape and porosity also influence the 
scattering phase function \citep[see e.g. Fig.~A2 in][]{pawellek-et-al-2023}, which along with disk viewing geometry determines its 
scattered light brightness; e.g., the surface brightness of a face-on disk made of highly forward-scattering grains would be lower 
than one with high inclination.

For now there are few constraints on exozodi dust composition, although it might be expected that this would have a strong dependence 
on exozodi origin, e.g., with dust created in situ having a lower volatile content than that originating further out in the system. 
Where mid-IR spectral features are seen in bright, warm exozodis, these can be a strong indicator for the dust origin 
\citep[e.g.,][]{Lisse2009}.

Finally, it is worth highlighting the importance of viewing orientation when assessing the surface brightness levels from 
circumstellar dust that could present confusion for exo-Earth imaging. For example, combining dust dynamics and expectations for 
near-infrared scattered light and mid-infrared thermal emission, \cite{Stark2015} found that in edge-on discs cold dust migrating 
inwards due to P-R drag might generate a ``pseudo-zodi'' mimicking the surface brightness of exozodiacal dust.

\textbf{Key Finding 5:} The surface brightness of exozodis and its dependence on wavelength are strongly influenced by the dust size 
distribution and composition.
One consequence is that the conversion from mid-IR flux to scattered light flux depends on exozodi origin.

\subsection{How common are different exozodi levels?}
\label{ss:howcommon}

One of the most pressing questions from the practicalities of designing a telescope capable of detecting an exo-Earth is the 
distribution of expected exozodi levels to be found around nearby stars. This is because exozodis present a noise source against which 
the planet's light needs to be distinguished, and this leads to requirements being set on the size and design of the telescope to 
ensure that this noise level is overcome \citep{Stark2015b, Kammerer2022, Quanz2022}. A typical rule of thumb which may be helpful is 
that exozodis that are $\ll 10$ times as bright as the Solar system's zodiacal cloud do not pose a significant constraint on exo-Earth 
imaging, whereas those $\gg 10$ times zodiacal cloud levels do.

It is worth acknowledging at the outset that the question in the title of this subsection is not well posed, since we need to define 
what we mean by {\it exozodi level}.
As elsewhere in this review we define this level in units of \textit{zodi}, where 1 zodi is the 
level of the Solar system’s zodiacal cloud \citep[in terms of surface density in the habitable zone, see][for the practical 
implementation of this unit]{Kennedy2015_LBTI}.
The zodi is a useful unit given its common usage, but we remind the reader of the caveats in \S \ref{ss:sizespatial},
i.e., that an exact replica of the zodiacal cloud would need to have the same size and spatial distribution,
and it is to be expected that the architecture of our zodiacal cloud is unique to our system, with its cometary source 
being strongly influenced by the planetary system architecture.

\begin{figure}[]
    \centering
     \includegraphics[width=0.45\textwidth]{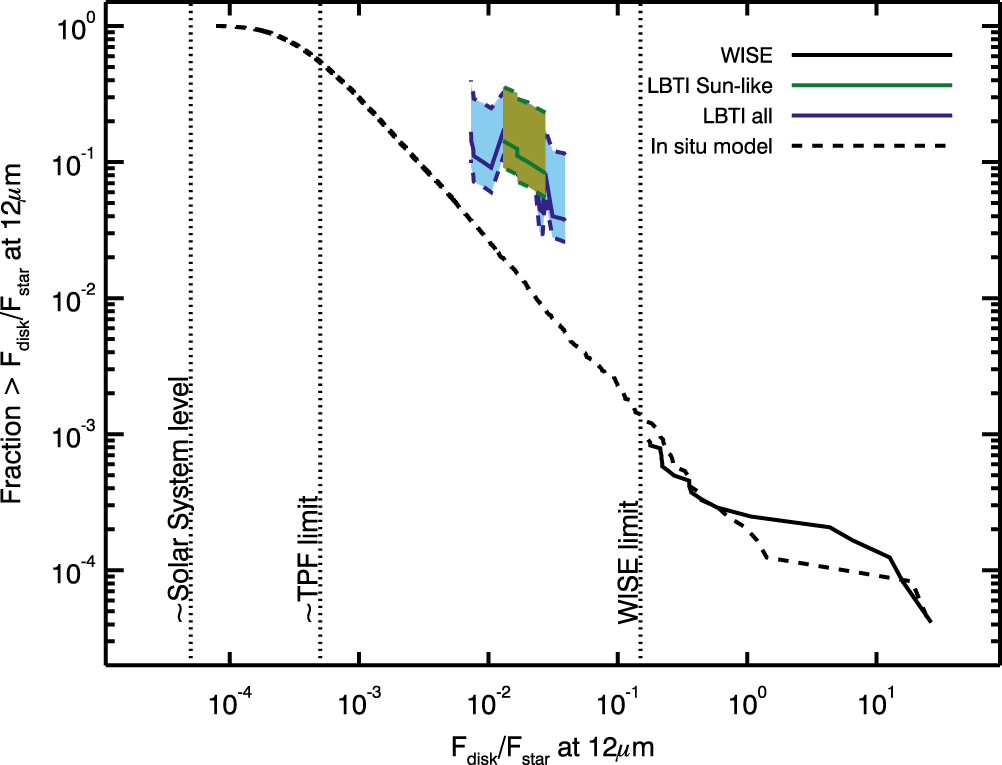}
    \caption{Exozodi luminosity function, i.e., the fraction of stars with fractional excesses above a given level
    $R_{12}=F_{\rm disk}/F_\star$ at 12$\mu$m \citep[$\copyright$ AAS, reproduced with permission from][]{Ertel2018}.
    The solid lines are observed fractions from WISE and LBTI, while the dashed line is a simple population model connected
    to the brightest excesses \citep[as described in][]{Kennedy2013}.}
    \label{fig:elf}
\end{figure}

When detecting an exozodi it is unlikely that we have access to the size and spatial distribution of its dust. Measurements are often 
made at a single wavelength, and the information provided is the total flux rather than its distribution. There is thus no guarantee 
that the measured flux provides an accurate assessment of dust levels in the habitable zone. Since habitable zone dust emits most 
efficiently in the mid-IR, surveys to characterise the frequency of exozodis are typically performed in the mid-IR. The exozodi 
luminosity function is the fraction of stars with 12$\mu$m dust emission above a given level relative to the star and is shown in 
Fig.~\ref{fig:elf}. This has been well characterised by WISE photometry for bright exozodis ($R_{12}>0.1$), which are rare being found 
around $\sim 0.1$\% of Sun-like stars \citep{Kennedy2013}. Fainter dust levels have been probed by nulling mid-IR interferometry with 
the HOSTS survey finding fractional excesses of $R_{12}>0.01$ around $\sim 20$\% of stars \citep{Ertel2020}. However, this means that 
thus far we can only detect and measure the frequency of exozodis at the $\sim 100$ zodi level.

To understand how common lower exozodi levels are requires an extrapolation of the exozodi luminosity function. One example of such an 
extrapolation is shown in Fig.~\ref{fig:elf} with a dashed line. This is the distribution that would be expected if the bright 
exozodis detected by WISE are transient phenomena, with their brightness decaying inversely with time since their appearance (e.g., 
due to collisional erosion). In such a model, for every bright exozodi detected we would expect many more stars to host fainter 
examples of the same phenomenon whose bright phase occurred at some point in the past. This is one example of a physically based model 
that can be used to make a prediction about the distribution down to fainter exozodi levels. However, it also illustrates how such a 
prediction requires an understanding of the origin of the exozodi. For example, a steady state origin for exozodis would make the 
extrapolation down from bright levels more complicated, since in that case a star's exozodi level is a property of its system that 
cannot readily be predicted from observations of other stars.

The correlation of exozodi detections in the HOSTS survey with the presence of an outer Kuiper belt-like disk provides one avenue to 
achieving an extrapolation down to fainter levels. This is because the population of outer Kuiper belts is relatively well 
characterised from far-IR observations \citep[e.g.,][]{Wyatt2007, Sibthorpe2018}. With a suitable model for how these belts populate 
the inner regions with dust \citep[e.g., through P-R drag or cometary scattering][]{Rigley2020}, the same population models could be 
used to make a prediction for the distribution of exozodi levels. While such population models would inevitably inherit any 
uncertainties remaining in our knowledge of the outer Kuiper belt population, as well as those in the models for how material is 
transported to the exozodi, they would be an improvement over current models which simply assume a log-normal distribution of exozodi 
levels that is characterised by its median zodi level \citep{Mennesson2014, Quanz2022}, and could be refined in line with our growing 
understanding of exozodi origins.

\textbf{Key Finding 6:} Theoretical (population) models can make predictions for the exozodi distribution, and point the way to 
methods to predict exozodi levels in a system.

\subsection{What information is needed to predict exozodi levels in a system?}

When identifying possible targets for exo-Earth imaging missions, or when assessing the likely yield of such missions, the approach is 
usually to assume that target stars’ exozodi levels are drawn randomly from some distribution, which is often taken to be a log-normal 
distribution with the median determined from a survey like HOSTS. Notwithstanding the discussion in \S \ref{ss:howcommon} which implies 
that a log-normal distribution may not be the best representation of the true distribution, this approach may be reasonable when 
considering the yield after having observed many stars, since then the exact exozodi level for any individual star is not important, 
only the distribution.

This approach is not, however, optimal when considering whether a specific star should be included in the target list for such a 
mission. For example, the HOSTS survey has already showed one way in which we can improve our estimate of the exozodi levels for 
specific stars: there is a correlation between the presence of cold outer dust and a detectable warm exozodi, and so we can use two 
different exozodi distributions according to whether or not the system is known to host cold dust.

Yet our understanding of the theory of exozodi origins shows that we can expect the presence of cold dust to be just one factor among 
many that determines the exozodi level. For example, if exozodi levels are set by inward transport of dust from the outer regions then 
the architecture of the planetary system will be a key determinant, since such planets can either eject dust or comets before it 
reaches the inner regions, or promote the transport of comets into those same regions. This illustrates how it is not only the 
presence of known planets that is important, but also their absence. For example, a star for which Saturn-mass planets in the $>5$\,au 
region have been excluded may be more likely to have a high exozodi level, both because of the absence of ejecting planets 
\citep{Bonsor2018} and the possibility of a chain of lower mass planets that can act as a comet conveyor belt \citep{Marino2018}.

A similar argument applies to the rather crude {\it presence of cold dust} criterion in the paragraph above, since the meaning of the 
absence of a detection of cold dust emission towards a particular star depends on the wavelengths at which it has been observed (e.g., 
by determining the temperature at which dust could have been detected) and to what depth \citep[e.g.,][]{Wyatt2008}. For some stars 
this may rule out cold dust down to levels approaching that of the Kuiper belt, while for others the constraint will be orders of 
magnitude higher, and this along with constraints on their planets would contribute to any predictions that can be made for what we 
might expect the level of dust to be in the stars' habitable zones. Similarly, while we may not yet know of any correlation between 
hot dust and the presence of dust in the habitable zone, the hot dust origin models predict that there should be such a correlation, 
and so any prediction for a star's exozodi level should take into account the presence or absence of dust at all temperatures.

Since both the star's planetary system and its cold dust levels are known to have some dependence on the star's properties, these 
should also be taken into account. For example, both the mass and radius and cold dust component are known to depend on stellar 
luminosity \citep{Greaves2003, Matra2018}, as is the architecture of its planetary system \citep{Bowler2016}, which also has a 
dependence on the star's metallicity and motion through the galaxy \citep{Fischer2005, Winter2020}. Since planetesimal populations are 
expected to decay with time, due to ejections or collisional erosion, stellar age is also an important factor. All-in-all, there are 
also many key stellar properties which may be expected to influence a system's likely exozodi level.

Not only can we be sure that all of the above factors are important in determining the exozodi level in a specific system, most of 
them will also be factors that determine the likelihood of the presence of a planet in the habitable zone and/or will have influenced 
its habitability. While it may not yet be clear how we can use knowledge of these factors to make a better prediction of exozodi 
levels, our understanding of exoplanetary systems will be vastly improved over the next decade or more as we build the telescopes that 
will eventually probe exo-Earths. For now it should already be apparent that the models used to predict the distribution of exozodis 
that were discussed in \S \ref{ss:howcommon} could also be used to predict how those distributions might depend on a system's other 
properties.

\textbf{Key Finding 7:} Models predict that exozodi levels will depend on many factors, including the properties of the star and its 
planets and disk, as well as constraints on the absence of such components.
This information should be used when assessing whether the star would be a suitable candidate for exo-Earth imaging.

\subsection{What is the connection between hot and warm exozodis?}

There seems to be no strong correlation between the presence of detectable warm and hot exozodis in individual systems 
\citep{Ertel2020}. However, hot-exozodical dust near stars should be quickly depleted through various physical processes (e.g. 
\citealt{Lebreton2013, Pearce2020}), and it cannot be sustained through collisions in an \textit{in-situ} planetesimal belt because 
such a belt would quickly collisionally erode \citep{Wyatt2007}. This leads to the currently favoured hypothesis that hot dust 
originates further out in systems, then somehow travels inwards to the hot-emission region. Such dust would therefore originate in, or 
travel through, the habitable zone where warm exozodis are located. Thus some connection between warm and hot exozodis could be 
expected. The lack of any detected correlation potentially tells us something about the sources, transport mechanisms and dynamics of 
both warm and hot dust.

One of the simplest explanations for hot exozodis, that P-R drag causes dust to migrate inwards from a distant planetesimal belt, can 
be discounted because it would produce too much mid-infrared emission to be compatible with observations \citep{vanLieshout2014, 
Sezestre2019}. Hence the presence of hot exozodis in systems without warm exozodis is interpreted as evidence that either dust is 
directly deposited in the hot-emission region by star-grazing comets \citep{Sezestre2019, Pearce2022Comets}, or that some unknown 
mechanism boosts the population of hot dust relative to warm dust \citep{Rieke2016, Kimura2020, Pearce2020}. The former case would 
imply that star-grazing comets pass through the habitable zone without releasing significant quantities of dust, in contrast to the 
Solar System's zodiacal cloud which is thought to be sustained via cometary fragmentation \citep{Rigley2022}. However, since 
star-grazing comets alone struggle to reproduce hot-exozodi observations \citep{Pearce2022Comets}, the currently favoured explanation 
is that some mechanism close to the star boosts the hot-dust population, potentially by trapping grains for long timescales. An 
efficient trap could sustain a hot-dust population with minimal dust inflow from comets or P-R drag, meaning that a tenuous dust 
source in the warm-emission region cannot be discounted as the origin of hot exozodis.

Similarly, the presence of warm exozodis in systems without hot exozodis suggests that either warm dust is unable to migrate inwards 
to the hot-emission region, possibly because it collisionally depletes or interacts with intervening planets, or that the hypothesised 
hot-dust-trapping mechanism does not operate in those systems.

Unfortunately, the lack of observational correlations between warm and hot exozodis, and the lack of understanding about how hot 
exozodis are supplied and sustained, means that we cannot currently use the presence or brightness of a system's hot exozodi to infer 
the properties of a warm exozodi, or \textit{vice versa}. If near-infrared excesses really are hot dust then warm and hot exozodis 
should be connected in some way, since the hot dust must almost certainly travel inward through the warm-emission region, but there 
are clearly other variables at play which we do not yet understand.
This means that, with current knowledge, it is not possible to say whether systems with hot exozodis would
be more or less favourable for future detection and characterisation of exo-Earths than other systems.

\textbf{Key Finding 8:} It would be helpful to know the connection between hot and warm exozodis!

\section{Conclusions}
\label{s:conclusion}

Exozodis are a common feature of planetary systems around main-sequence stars.
They are currently the subject of increased interest, because they could potentially impede future attempts to
image habitable-zone exoplanets.
This review summarises our theoretical understanding of exozodis, based on models that attempt to reproduce exozodi observations.
The current view is that exozodis are comprised of dust grains, either located in the habitable zone or closer to the star.
This dust is thought to originate from further out in the system, and somehow get transported inwards.
Possible explanations include dust released by fragmenting comets, and dust migrating inwards due to radiation forces.
However, the origin, distribution, composition and dynamics of this dust remain uncertain, which makes it difficult to assess
its potential impact on exoplanet imaging.
This review highlights several key questions which must be answered if we are to assess the impact of exozodis on exoplanet
imaging, and the contribution of exozodi theory to answering them.
These include how the dust is delivered, how can we predict exozodi levels from system properties, what features planets would
impart in dust, and the effect of composition on exozodi observability.
Theoretical work in the near future should focus on answering these questions, partly to further our understanding of
planetary systems, but also to ascertain how exozodis might impact exoplanet imaging.



\begin{acknowledgments}
MCW and JKR acknowledge support from the Science and Technology Facilities Council (STFC)
grant number ST/W000997/1.
TDP is supported by a UKRI Stephen Hawking Fellowship and a Warwick Prize Fellowship, the latter made possible by a
generous philanthropic donation.
VFG acknowledges funding from the National Aeronautics and Space Administration through the Exoplanet Research Program under
Grants No. 80NSSC21K0394 and 80NSSC23K1473 (PI: S. Ertel), and Grant No 80NSSC23K0288 (PI: V. Faramaz).
\end{acknowledgments}


\bibliography{bib}{}
\bibliographystyle{aasjournalv7}


\end{document}